\documentclass[runningheads]{llncs}
\usepackage{graphicx}
\usepackage{subcaption}
\usepackage{cite}
\usepackage{acro}
\usepackage{float}
\usepackage{xcolor}
\usepackage{amsmath}
\usepackage{multirow}
\usepackage{multicol}
\usepackage{wrapfig}
\usepackage{booktabs}   
\usepackage{subcaption}
\usepackage{wrapfig}
\usepackage{float}
\usepackage[english]{babel}
\usepackage{hyperref}
\usepackage{adjustbox}
\usepackage{threeparttable}
\usepackage{utfsym}

\begin{document}
\title{Longitudinal Multimodal Transformer Integrating Imaging and Latent Clinical Signatures From Routine EHRs for Pulmonary Nodule Classification}
\titlerunning{}


\author{Thomas Z. Li\inst{1}, John M. Still\inst{1}, Kaiwen Xu\inst{1}, Ho Hin Lee\inst{1}, Leon Y. Cai\inst{1}, Aravind R. Krishnan\inst{1}, Riqiang Gao\inst{2}, Mirza S. Khan\inst{3}, Sanja Antic\inst{4}, Michael Kammer\inst{4}, Kim L. Sandler\inst{4}, Fabien Maldonado\inst{4}, Bennett A. Landman\inst{1}, Thomas A. Lasko\inst{1}}

\institute{Vanderbilt University, Nashville, TN 37212, USA \and
Digital Technology and Innovation, Siemens Healthineers, Princeton NJ 08540, USA \and
Saint Luke’s Mid America Heart Institute, Kansas City, MO 64111, USA\and
Vanderbilt University Medical Center, Nashville, TN 37235, USA}
\maketitle  
The accuracy of predictive models for solitary pulmonary nodule (SPN) diagnosis can be greatly increased by incorporating repeat imaging and medical context, such as electronic health records (EHRs). However, clinically routine modalities such as imaging and diagnostic codes can be asynchronous and irregularly sampled over different time scales which are obstacles to longitudinal multimodal learning. In this work, we propose a transformer-based multimodal strategy to integrate repeat imaging with longitudinal clinical signatures from routinely collected EHRs for SPN classification. We perform unsupervised disentanglement of latent clinical signatures and leverage time-distance scaled self-attention to jointly learn from clinical signatures expressions and chest computed tomography (CT) scans. Our classifier is pretrained on 2,668 scans from a public dataset and 1,149 subjects with longitudinal chest CTs, billing codes, medications, and laboratory tests from EHRs of our home institution. Evaluation on 227 subjects with challenging SPNs revealed a significant AUC improvement over a longitudinal multimodal baseline (0.824 vs 0.752 AUC), as well as improvements over a single cross-section multimodal scenario (0.809 AUC) and a longitudinal imaging-only scenario (0.741 AUC). This work demonstrates significant advantages with a novel approach for co-learning longitudinal imaging and non-imaging phenotypes with transformers. Code available at \href{https://github.com/MASILab/lmsignatures}{https://github.com/MASILab/lmsignatures}.
\begin{abstract}

\keywords{Multimodal Transformers, Latent Clinical Signatures, Pulmonary Nodule Classification.}
\end{abstract}
\section{Introduction}
The absence of highly accurate and noninvasive diagnostics for risk-stratifying benign vs malignant solitary pulmonary nodules (SPNs) leads to increased anxiety, costs, complications, and mortality~\cite{massion2014,Rivera2013}. The use of noninvasive methods to discriminate malignant from benign SPNs is a high-priority public health initiative~\cite{gould2015,gomez-Saez2014}. Deep learning approaches have shown promise in classifying SPNs from longitudinal chest computed tomography (CT)~\cite{ardila2019, Gao2020, liao2019evaluate, huang2019prediction}, but approaches that only consider imaging are fundamentally limited. Multimodal models generally outperform single modality models in disease diagnosis and prediction~\cite{mohsen2022artificial}, and this is especially true in lung cancer which is heavily contextualized through non-imaging risk factors~\cite{mcwilliams2013probability, gao2021deep, vanguri2022multimodal}. Taken together, these findings suggest that learning across both time and multiple modalities is important in biomedical predictive modeling, especially SPN diagnosis. However, such an approach that scales across longitudinal multimodal data from comprehensive representations of the clinical routine has yet to be demonstrated~\cite{mohsen2022artificial}.

\textbf{Related work.} Directly learning from routinely collected electronic health records (EHRs) is challenging because observations within and between modalities can be sparse and irregularly sampled. Previous studies overcome these challenges by aggregating over visits and binning time series within a Bidirectional Encoder Representations from Transformers (BERT) architecture~\cite{choi2016retain, labach2023effective, rasmy2021med, li2020behrt}, limiting their scope to data collected on similar time scales, such as ICU measurements,~\cite{tipirneni2022self,clinicalbert}, or leveraging graph guided transformers to handle asynchrony~\cite{zhang2021graph}. Self-attention ~\cite{vaswani2017attention} has become the dominant technique for learning powerful representations of EHRs with trade-offs in interpretability and quadratic scaling with the number of visits or bins, which can be inefficient with data spanning multiple years. In contrast, others address the episodic nature of EHRs by converting non-imaging variables to continuous longitudinal curves that provide the instantaneous value of categorical variables as intensity functions~\cite{lasko2014efficient} or continuous variables as latent functions~\cite{lasko2015nonstationary}. Operating with the hypothesis that distinct disease mechanisms manifest independently of one another in a probabilistic manner, one can learn a transformation that disentangles latent sources, or clinical signatures, from these longitudinal curves. Clinical signatures learned in this way are expert-interpretable and have been well-validated to reflect known pathophysiology across many diseases~\cite{lasko2019computational,lasko2023ehr}. Given that several clinical risk factors have been shown to independently contribute to lung cancer risk, these signatures are well poised for this predictive task. Despite the wealth of studies seeking to learn comprehensive representations of routine EHRs, these techniques have not been combined with longitudinal imaging. 

\textbf{Present work.} In this work, we jointly learn from longitudinal medical imaging, demographics, billing codes, medications, and lab values to classify SPNs. We converted 9195 non-imaging event streams from the EHR to longitudinal curves to impute cross-sections and synchronize across modalities. We use Independent Component Analyses (ICA) to disentangle latent clinical signatures from these curves, with the hypothesis that the disease mechanisms known to be important for SPN classification can also be captured with probabilistic independence. We leverage a transformer-based encoder to fuse features from both longitudinal imaging and clinical signature expressions sampled at intervals ranging from weeks to up to five years. Due to the importance of time dynamics in SPN classification, we use the time interval between samples to scale self-attention with the intuition that recent observations are more important to attend to than older observations. Compared with imaging-only and a baseline that aggregates longitudinal data into bins, our approach allowed us to incorporate additional modalities from routinely collected EHRs, which led to improved SPN classification. 

\begin{figure}[t!]
\centering
\includegraphics[width=\textwidth]{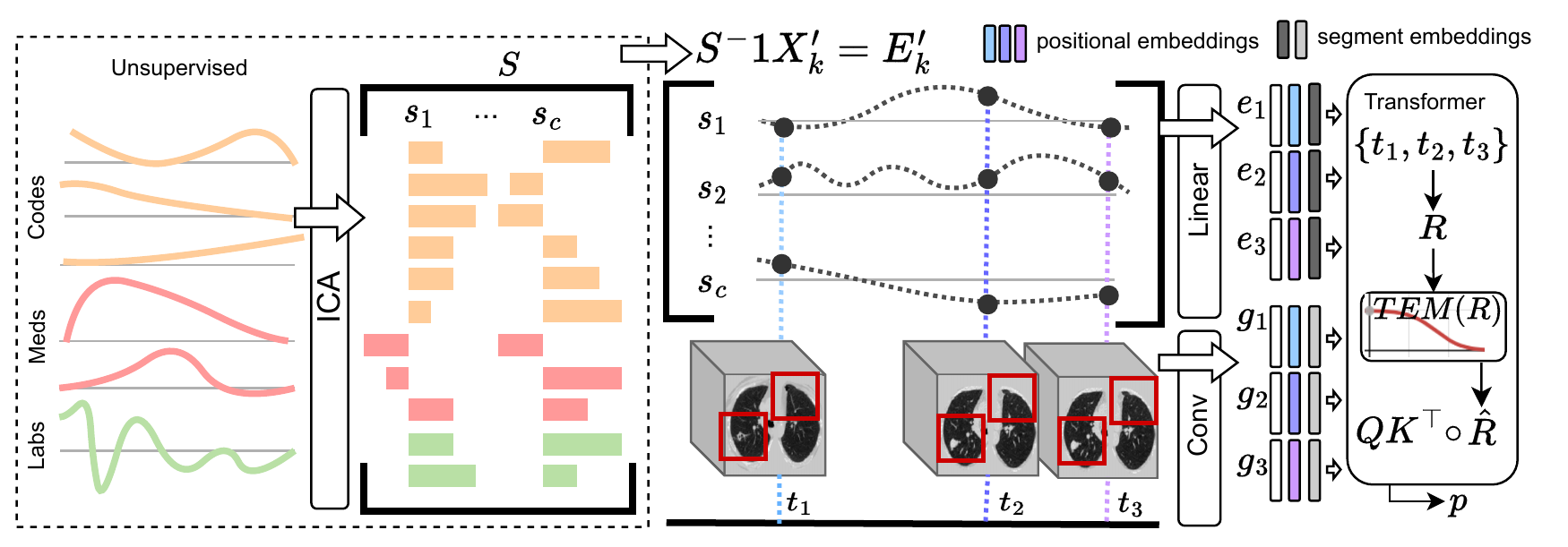}
\caption{Left: Event streams for non-imaging variables are transformed into longitudinal curves. ICA learns independent latent signatures, $S$, in an unsupervised manner on a large non-imaging cohort. Right: Subject $k$'s expressions of the signatures, $E'_k$, are sampled at scan dates. Input embeddings are the sum of 1) token embedding derived from signatures or imaging, 2) a fixed positional embedding indicating the token's position in the sequence, and 3) a learnable segment embedding indicating imaging or non-imaging modality. The time interval between scans is used to compute a time-distance scaled self-attention. This is a flexible approach that handles asynchronous modalities, incompleteness over  varying sequence lengths, and irregular time intervals.}
\end{figure}

\section{Methods}
\textbf{Latent clinical signatures via probabilistic independence}. We obtained event streams for billing codes, medications, and laboratory tests across the full record of each subject in our EHR cohorts (up to 22 years). After removing variables with less than 1000 events and mapping billing codes to the SNOMED-CT ontology ~\cite{pmid33496673}, we arrived at 9195 unique variables. We transformed each variable to a longitudinal curve at daily resolution, estimating the variable’s instantaneous value for each day~\cite{lasko2019computational}. We used smooth interpolation for continuous variables~\cite{fritsch1984method} or a continuous estimate of event density per time for event data. Previous work used Gaussian process inference to compute both types of curves~\cite{lasko2015nonstationary,lasko2014efficient}. For this work we traded approximation for computational efficiency. To encode a limited memory into the curve values, each curve was smoothed using a rolling uniform mean of the past 365 days (Fig. 1, left). 

We use an ICA model to estimate a linear decomposition of the observed curves from the EHR-Pulmonary cohort to independent latent sources, or clinical signatures. Formally, we have dataset $D_\text{EHR-Pulmonary} = \{L_k \mid k=1,\ldots,n\}$ with longitudinal curves denoted as $L_k = \{l_i | i=1,\ldots,9195\}$. We randomly sample $l_i\;\forall i \in [1,9195]$ at a three-year resolution and concatenate samples across all subjects as $x_i \in R^m$. For $D_\text{EHR-Pulmonary}$, $m$ was empirically found to be $630037$. We make a simplifying assumption that $x_i$ is a linear mixture of $c$ latent sources, $s$, with longitudinal expression levels $e \in R^m$:
\begin{equation}
    x_i = e_1 s_{i,1} + e_2 s_{i,2} + \ldots + e_c s_{i, c}
\end{equation}
The linear mixture is then $X = SE$ with $x_i$ forming the rows of X, $S \in R^{9195\times c}$ denoting the independent latent sources and $E \in R^{c\times m}$ denoting the expression matrix. We set $c=2000$ and estimated $S$ in an unsupervised manner using FastICA~\cite{hyvarinen1999fast}. Given longitudinal curves for another cohort, for instance $D_\text{Image-EHR} = \{X_k' \mid k=1, \ldots, n\}$, we obtain expressions of clinical signatures for subject $k$ via $E_k' = S^{-1}X_k'$ (Fig. 1, left).

\textbf{Longitudinal multimodal transformer (TDSig)}. We represent our multimodal datasets $D_\text{Image-EHR}$ and $D_\text{Image-EHR-SPN} = \{(E_k, G_k) \mid k=1, \ldots, n\}$ as a sequence of clinical expressions $E_k=\{e_{k,1},\ldots,e_{k,T}\}$ sampled at the same dates as images $G_k=\{g_{k,1},\ldots, g_{k,T}\}$, where $T$ is the maximum sequence length. We set $T=3$ and added a fixed padding embedding to represent missing items in the sequence. Embeddings that incorporate positional and segment information are computed for each item in the sequence (Fig. 1, right). Token embeddings for images are a convolutional embeddings of five concatenated 3D patches proposed by a pretrained SPN detection model ~\cite{liao2019evaluate}. We use a 16-layer ResNet ~\cite{he2015deep} to compute this embedding. Likewise, token embeddings for clinical signature expressions are linear transformations to the same dimension as imaging token embeddings. The sequence of embeddings are then passed through a multi-headed Transformer. All embeddings except the nodule detection model are co-optimized with the Transformer. We will refer to this approach as TDSig.

\textbf{Time-distance self-attention}. Following ~\cite{Gao2020,li2022time,wu2020transformer}, we intuit that if medical data is sampled as a cross-sectional manifestation of a continuously progressing phenotype, we can use a temporal emphasis model (TEM) emphasize the importance of recent observations over older ones. Additionally, self-attention is masked for padded embeddings, allowing our approach to scale with varying sequence lengths across subjects. Formally, if subject $k$ has a sequence of $T$ images at relative acquisition days $t_1\ldots t_T$, we construct a matrix $R$ of relative times with entries $R_{i,j} = |t_T – t_i|$ where $t_i$ is the acquisition day of tokens $\hat{e}_{k,i}$ and $\hat{g}_{k,i}$, or 0 if they are padded embeddings. We map the relative times in $R$ to a [0,1] value in $\hat{R}$ using a TEM of the form:
\begin{equation}
    \hat{R}_{i,j}=\text{TEM}(R_{i,j}) = \frac{1}{1+ exp(bR_{i,j} - c)}
\end{equation}
This is a flipped sigmoid function that monotonically decreases with the relative time from the most recent observation. Its slope of decline and decline offset are governed by learnable non-negative parameters $b$ and $c$ respectively. A separate TEM is instantiated for each attention head, with the rationale that separate attention heads can learn to condition on time differently. The transformer encoder computes query, key, and value matrices as linear transformations of input embedding $H=\{\hat{E}\mathbin\Vert \hat{G}\}$ at attention head $p$
\begin{equation*}
    Q_p = H_pW_p^Q \qquad K_p = H_pW_p^K \qquad V_p = H_pW_p^V
\end{equation*}
TEM-scaled self-attention is computed via element-wise multiplication of the query-key product and $\hat{R}$:
\begin{equation}
    \text{softmax}\left(\frac{\text{ReLU}(Q_pK_p^\top + M)\circ \hat{R}}{\sqrt{d}}\right)V_p
\end{equation}
where M is the padding mask~\cite{vaswani2017attention} and $d$ is the dimension of the query and key matrices. ReLU gating of the query-key product allows the TEM to adjust the attention weights in an unsigned direction. 

\textbf{Baselines}
We compared against a popular multimodal strategy that aggregates event streams into a sequence of bins as opposed to our method of extracting instantaneous cross-sectional representations. For each scan, we computed a TF-IDF~\cite{schutze2008introduction} weighted vector from all billing codes occurring up to one year before the scan acquisition date. We passed this through a published Word2Vec-based medical concept embedding~\cite{finch2021exploiting} to compute a contextual representation $\in R^{100}$. This, along with the subject's scans, formed a sequence that was used as input to a model we call TDCode2vec. Our search for contextual embeddings for medications and laboratory values did not yield any robust published models that were compatible with our EHR's nomenclature, so these were not included in TDCode2vec.  We also performed experiments using only image sequences as input, which we call TDImage. Finally, we implemented single cross-sectional versions of TDImage, TDCode2vec, and TDSig, CSImage, CSCode2vec, and CSSig respectively, using the scan date closest to the lung malignancy diagnosis for cases or SPN date for controls. All baselines except CSImage, which employed a multi-layer perceptron directly after the convolutional embedding, used the same architecture and time-distance self-attention as TDSig. The transformer encoders in this study were standardized to 4 heads, 4 blocks, input token size of 320, multi-layer perception size of 124, self-attention weights of size 64. This work was supported by Pytorch 1.13.1, CUDA 11.7.

\begin{table*}[t!]
\small
    \centering
        \caption{Breakdown of modalities, size, and longitudinality of each dataset.}
            \begin{threeparttable}[b]
                \begin{tabular}{*{1}{l}*{5}{c}*{2}{c}}
                    \toprule
                    & \multicolumn{5}{c}{Modalities} &  \multicolumn{2}{c}{Counts (cases/controls)} \\
                    \midrule
                    & Demo & Img & Code & Med & Lab & Subjects & Scans \\
                    \midrule
                    EHR-Pulmonary & \usym{1F5F8} & - & \usym{1F5F8}& \usym{1F5F8} & \usym{1F5F8} & 288,428 & - \\
                    NLST & \usym{1F5F8} &\usym{1F5F8} & - & - & - & 533/801 & 1066/1602 \\
                    Image-EHR & \usym{1F5F8} &\usym{1F5F8} & \usym{1F5F8} & \usym{1F5F8} & \usym{1F5F8} & 257/665 & 641/1624 \\
                    Image-EHR-SPN & \usym{1F5F8} &\usym{1F5F8} & \usym{1F5F8} & \usym{1F5F8} & \usym{1F5F8} & 58/169 & 76/405 \\
                    \bottomrule

                \end{tabular}
                \begin{tablenotes}
                    \item Demo: Demographics, Img: Chest CTs, Code: ICD billing codes, Med: Medications, Lab: Laboratory tests.
                \end{tablenotes}
            \end{threeparttable}
\end{table*}
\section{Experimental Setup}
\textbf{Datasets}. This study used an imaging-only cohort from the NLST~\cite{national2011reduced} and three multimodal cohorts from our home institution with IRB approval (Table 1). For the \textbf{NLST} cohort (https://cdas.cancer.gov/nlst/), we identified cases who had a biopsy-confirmed diagnosis of lung malignancy and controls who had a positive screening result for an SPN but no lung malignancy. We randomly sampled from the control group to obtain a 4:6 case control ratio. Next, \textbf{EHR-Pulmonary} was the unlabeled dataset used to learn clinical signatures in an unsupervised manner. We searched all records in our EHR archives for patients who had billing codes from a broad set of pulmonary conditions, intending to capture pulmonary conditions beyond just malignancy. Additionally, \textbf{Image-EHR} was a labeled dataset with paired imaging and EHRs. We searched our institution’s imaging archive for patients with three chest CTs within five years. In the EHR-Image cohort, malignant cases were labeled as those with a billing code for lung malignancy and no cancer of any type prior. Importantly, this case criteria includes metastasis from cancer in non-lung locations. Benign controls were those who did not meet this criterion. Finally, \textbf{Image-EHR-SPN} was a subset of Image-EHR with the inclusion criteria that subjects had a billing code for an SPN and no cancer of any type prior to the SPN. We labeled malignant cases as those with a lung malignancy billing code occurring within three years after any scan and only used data collected before the lung malignancy code. All data within the five-year period were used for controls. We removed all billing codes relating to lung malignancy. A description of the billing codes used to define SPN and lung cancer events are provided in Supplementary 1.2.

\textbf{Training and Validation}
All models were pretrained with the NLST cohort after which we froze the convolutional embedding layer. While this was the only pretraining step for image-only models (CSImage and TDimage), the multimodal models underwent another stage of pretraining using the Image-EHR cohort with subjects from Image-EHR-SPN subtracted. In this stage, we randomly selected one scan and the corresponding clinical signature expressions for each subject and each training epoch. Models were trained until the running mean over 100 global steps of the validation loss increased by more than 0.2. For evaluation, we performed five-fold cross-validation with Image-EHR-SPN, using up to three of the most recent scans in the longitudinal models. We report the mean AUC and 95\% confidence interval from 1000 bootstrapped samples, sampling with replacement from the pooled predictions across all test folds. A two-sided Wilcoxon signed-rank test was used to test if differences in mean AUC between models were significant.

\textbf{Reclassification Analysis}
We performed a reclassification analysis of low, medium, and high-risk tiers separated by thresholds of 0.05 and 0.65, which are the cutoffs used to guide clinical management. Given a baseline comparison, our approach reclassifies a subject correctly if it predicts a higher risk tier than the baseline in cases, or a lower risk tier than the baseline in controls (Fig. 2). 
\begin{figure}[t!]
\centering
\includegraphics[width=\textwidth]{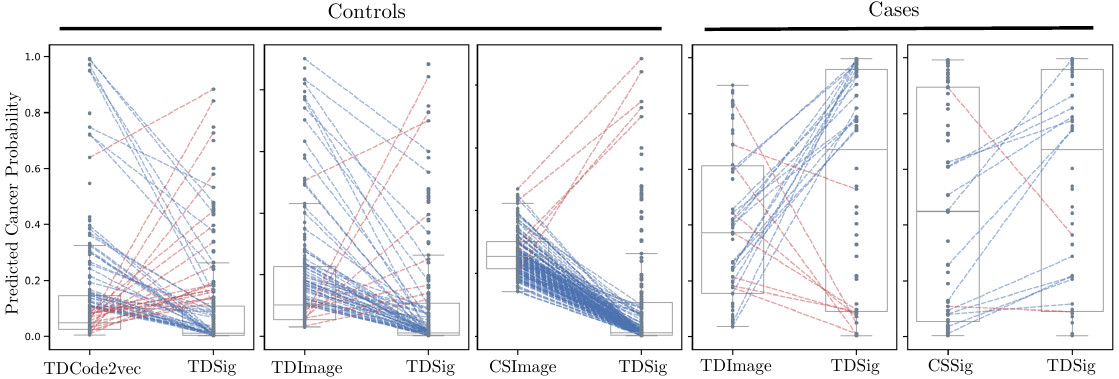}
\caption{A comparison of median and interquartile range of predicted probabilities reveals that TDSig is more correctly confident than baselines. Blue and red indicate subjects that were correctly and incorrectly reclassified by TDSig respectively. When compared to these baselines, TDSig is more often reclassifying correctly than not.}
\end{figure}

\begin{table*}[t!]
    \centering
        \caption{Performance on SPN classification using different approaches and modalities.}
        \begin{adjustbox}{width=\textwidth}
            
            \begin{threeparttable}[b]
                \begin{tabular}{*{1}{l}*{1}{c}*{5}{c}|*{2}{c}}

                    \toprule
                    \multicolumn{2}{c}{}& \multicolumn{5}{c|}{Modalities} & \multicolumn{2}{c}{Pretrain} \\
                    \midrule
                    \multicolumn{1}{c}{} & Mean AUC [95\% CI] & Img & Demo & Code & Med & \multicolumn{1}{c|}{Lab}& NLST & Image-EHR \\
                    \midrule
                    CSImage &  0.7392 [0.7367, 0.7416] & \usym{1F5F8}&- & - & - & - & \usym{1F5F8} & - \\
                    CSCode2vec &  0.7422 [0.7398, 0.7447] & \usym{1F5F8} & \usym{1F5F8}& \usym{1F5F8} & - & - & \usym{1F5F8} & \usym{1F5F8} \\
                    CSSig &  0.8097 [0.8075, 0.8120] & \usym{1F5F8} & \usym{1F5F8}&\usym{1F5F8} & \usym{1F5F8} & \usym{1F5F8} & \usym{1F5F8} & \usym{1F5F8} \\
                    TDImage &  0.7406 [0.7381, 0.7432] & \usym{1F5F8} &-& - & - & - & \usym{1F5F8} & - \\
                    TDCode2vec &  0.7524 [0.7499, 0.7550] & \usym{1F5F8} &\usym{1F5F8}& \usym{1F5F8} & - & - & \usym{1F5F8} & \usym{1F5F8} \\
                    \textbf{TDSig} & \textbf{0.8238 [0.8216, 0.8260]*} & \usym{1F5F8} & \usym{1F5F8}&\usym{1F5F8} & \usym{1F5F8} & \usym{1F5F8} & \usym{1F5F8} & \usym{1F5F8} \\
                    \bottomrule
                \end{tabular}

                \begin{tablenotes}
                    \item $\ast$:  $p<0.01$ against all other methods.
                \end{tablenotes}
            \end{threeparttable}
        \end{adjustbox}
\end{table*}

\begin{figure}[t!]
\centering
\includegraphics[width=\textwidth]{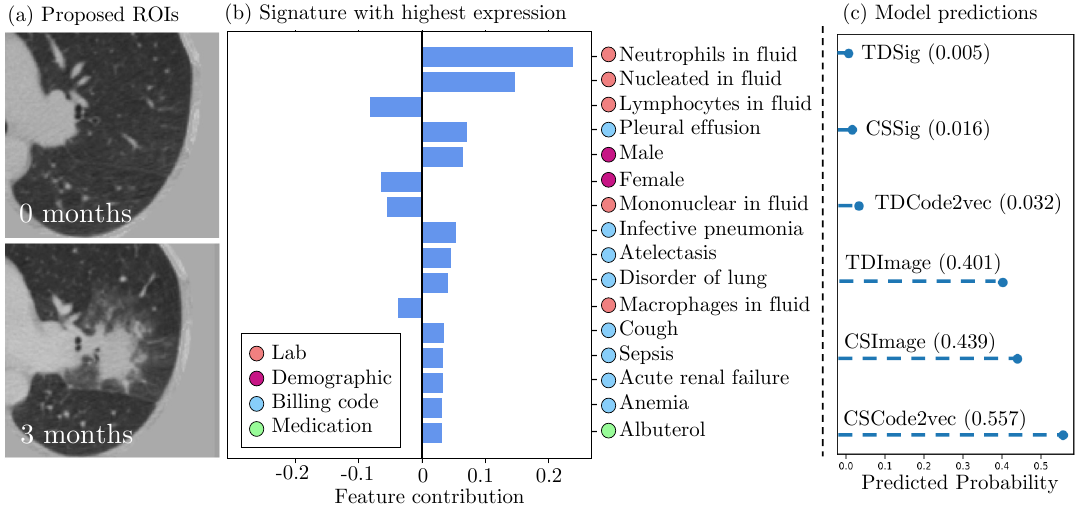}
\caption{This is a control subject who developed a lesion over 3 months (a), to which the imaging-only approaches assigned a cancer probability of 0.4 (c). However, the subject's highest expressed clinical signature at the 3-month mark was a new pattern of bacterial pneumonia (b), offering to the model a benign explanation of an image that it would otherwise be less correctly confident in.}
\end{figure}

\section{Results}
 The significant improvement with TDSig over CSSig demonstrates the advantage of longitudinally in the context of combining images and clinical signatures (Table 2). There were large performance gaps between TDSig and TDCode2vec, as well as between CSSig and CSCode2vec, demonstrating the advantage of clinical signatures over a binned embedding strategy. Cross-sectional embedded billing codes did not significantly improve performance over images alone (CSCode2vec vs CSImage, $p=0.56$), but adding clinical signatures did (CSSig vs CSImage, $p<0.01$; TDSig vs TDImage, $p<0.01$) and the greatest improvement in longitudinal data over single cross sections occurred when clinical signatures were included.
 
 For control subjects, TDSig correctly/incorrectly reclassified 40/18 from TDCode2vec, 54/8 from TDImage, 12/18 from CSSig, 104/7 from CSCode2vec, and 125/5 from CSImage. For case subjects, TDSig correctly/incorrectly reclassified 13/10 from TDCode2vec, 17/8 from TDImage, 12/2 from CSSig, 23/16 from CSCode2vec, and 29/16 from CSImage (Fig. 2). Full reclassification matrices are reported in Supplementary 6.1. On qualitative inspection of a control subject, clinical signatures likely added clarity to benign imaging findings that were difficult for baseline approaches to classify (Figure 3).

\section{Discussion and Conclusion}
This work presents a novel transformer-based strategy for integrating longitudinal imaging with interpretable clinical signatures learned from comprehensive multimodal EHRs. We demonstrated large performance gains in SPN classification compared with baselines, although calibration of our models is needed to assess clinical utility. We evaluated on clinically-billed SPNs, meaning that clinicians likely found these lesions difficult enough to conduct a clinical workup. In this setting, we found that adding clinical context increased the performance gap between longitudinal data and single cross-sections. Our clinical signatures incorporated longitudinality and additional modalities to build a better representation of clinical context than binned embeddings. We release our implementation at \href{https://github.com/MASILab/lmsignatures}{https://github.com/MASILab/lmsignatures}.

The lack of longitudinal multimodal datasets has long been a limiting factor ~\cite{mohsen2022artificial} in conducting studies such as ours. One of our contributions is demonstrating training strategies in a small-dataset, incomplete-data regime. We were able to overcome our small cohort size (Image-EHR-SPN) by leveraging unsupervised learning on datasets without imaging (EHR-Pulmonary), pretraining on public datasets without EHRs (NLST), and pretraining on paired multimodal data with noisy labels (Image-EHR) within a flexible transformer architecture.

Our approach of sampling cross-sections where clinical decisions are likely to be made scales well with long, multi-year observation windows, which may not be true for BERT-based embeddings~\cite{rasmy2021med, li2020behrt}. We did not compare against these contextual embeddings because none have been publically released, but integrating these with longitudinal imaging is an area of future investigation.

\section{Acknowledgements}
This research was funded by the NIH through R01CA253923-02 and in part by NSF CAREER 1452485 and NSF 2040462. This research is also supported by ViSE through T32EB021937-07 and the Vanderbilt Institute for Clinical and Translational Research through UL1TR002243-06. We thank the National Cancer Institute for providing data collected through the NLST.

%
%
\bibliographystyle{splncs04}
\bibliography{main_and_supp}

\newpage
\setcounter{section}{1}
\section*{Supplementary Material}
\subsection{Reclassification analysis}
\begin{figure}[!htb]
\centering
\includegraphics[width=\textwidth]{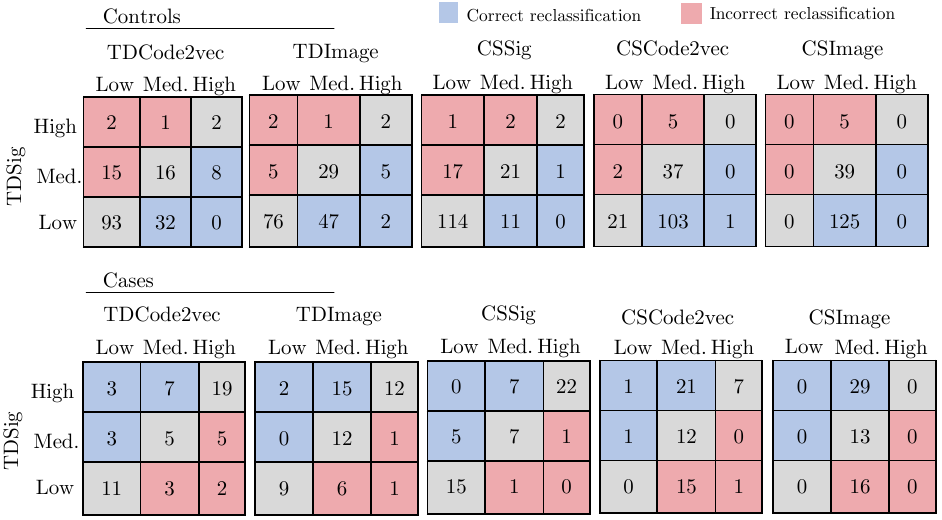}
\caption{Confusion matrices for TDSig reclassification from baseline approaches, separated by controls and cases.}
\end{figure}
\subsection{Dataset criteria}
\begin{table*}[t!]
\small
    \centering
        \caption{Billing codes used to curate Image-EHR and Image-EHR-SPN}
            \begin{threeparttable}[b]
                \begin{tabular}{*{3}{l}}
                    \toprule
                    \textbf{Version} & \multicolumn{1}{c}{\textbf{Code}} &  \textbf{Description}\\
                    \midrule
                    \multicolumn{3}{l}{\textbf{Phenotype: Solitary Pulmonary Nodule}} \\
                    ICD-9 & 793.11 & Solitary pulmonary nodule \\
                    ICD-10 & R91.1 & Solitary pulmonary nodule \\
                    \midrule
                    \multicolumn{3}{l}{\textbf{Phenotype: Lung cancer}} \\
                    ICD-9 & 162$^\dagger$ & Malignant neoplasm of trachea bronchus and lung \\
                    ICD-9 & 197.0 & Secondary malignant neoplasm of lung \\
                    ICD-9 & 209.21 & Malignant carcinoid tumor of the bronchus and lung \\
                    ICD-9 & 176.4 & Kaposi's sarcoma, lung \\
                    ICD-10 & C34 & Malignant neoplasm of bronchus and lung \\
                    ICD-10 & C7A.090 & Malignant carcinoid tumor of the bronchus and lung \\
                    ICD-10 & C46.5$^\ast$ & Kaposi's sarcoma of lung \\
                    ICD-10 & C78.0$^\ast$ & Secondary malignant neoplasm of lung \\
                    \bottomrule

                \end{tabular}
                \begin{tablenotes}
                    \item $^\dagger$Includes all sub-categories under this code in the ICD hierarchy except 162.0 "Malignant neoplasm of trachea". $^\ast$Includes all sub-categories under this code in the ICD hierarchy.
                \end{tablenotes}
            \end{threeparttable}
\end{table*}

\end{document}